\documentclass[prl,twocolumn,showpacs,superscriptaddress,floatfix]{revtex4}
\usepackage{graphicx,amsfonts,amssymb,amsmath}

\usepackage{hyperref}
\newif\ifhyper
\hypertrue
\ifhyper
\hypersetup{
   citecolor = {green},
   colorlinks = {true}, 
   urlcolor = {blue} 
}
\fi
\newcommand{\rmi}{\mathrm{i}}
\newcommand{\ket}[1]{|#1\rangle} 
\newcommand{\bra}[1]{\langle#1|} 
\newcommand{\braket}[2]{\langle#1|#2\rangle} 

\begin{document}
\title{Semi-classical Analysis of Spin Systems near Critical Energies}
\author{Pedro Ribeiro}
\affiliation{Laboratoire de Physique Th\'eorique de la Mati\`ere Condens\'ee, CNRS UMR 7600,
Universit\'e Pierre et Marie Curie, 4 Place Jussieu, 75252 Paris Cedex 05, France}
\author{Thierry Paul}
\affiliation{CNRS, DMA, Ecole Normale Sup\'erieure, 45 rue d'Ulm, 75230 Paris Cedex 05, France} 
\begin{abstract}
The spectral properties of $su(2)$ Hamiltonians are studied for energies near the critical classical energy $\varepsilon_c$ for which the corresponding classical dynamics presents hyperbolic points (HP). 
A general method leading to an algebraic relation for eigenvalues in the vicinity of $\varepsilon_c$ is obtained in the thermodynamic limit,  when the semi-classical parameter $n^{-1}=(2s)^{-1}$ goes to zero (where $s$ is the total spin of the system). Two applications of this method are given and compared with numerics.  
Matrix elements of observables, computed between states with energy near  $\varepsilon_c$, are also computed and shown to be in agreement with the numerical results.
\end{abstract}
\pacs{
03.65.Sq, 
05.30.-d, 
21.60.Ev 
}      
\maketitle
%
%
%
%
{\bf Introduction.}
Bohr-Sommerfeld (B-S) quantization formul\ae\ for nonregular values of the energy parameter have been set up in \cite{Connor_1969, Verdiere_1999} in the case of quantum Hamiltonians in the Schr\"odinger representation. They differ from the regular case and show a logarithmic accumulation of the spectrum near energies corresponding to hyperbolic fixed points.
In this letter we construct a general method to overcome the breakdown of standard B-S quantization near singular spectral points in the case of spins systems, a situation non covered by earlier results. 

$su(2)$ Hamiltonians arise naturally in many areas of physics, in the study of mutually interacting spins $1/2$ systems or due to symmetries present in collective bosonic Hamiltonians. 
For such models the analog of the semi-classical limit is obtained when the number of interacting sub-systems increases (thermodynamic limit). The semi-classical parameter $\hbar$ is replaced by the  the inverse of number of interacting sub-systems $n^{-1}$ which is related to the size of the  considered 
$su(2)$ representation, $n = 2 s$.
A typical example is given by the Lipkin-Meshkov-Glick (LMG) model proposed in 1965 to describe shape phase transition in nuclei \cite{Lipkin_1965}. This model is used to describe magnetic properties of molecules \cite{Garanin_1998}, interacting bosons in double-well structures \cite{ Ulyanov_1992} and to investigate the role of entanglement in quantum phase transitions (QPT) \cite{Vidal_2004_b}.

QPT \cite{Sachdev_1999} arising at zero temperature are related to non-analyticities of the ground state as a function of the Hamiltonian coupling constants. Since the non-analyticities involved are generically algebraic, this kind of phase transitions are characterized by a set of critical exponents describing how physical quantities (density of states, excitation gap, observables) behave in the vicinity of such points. Recently, non-analyticities arising within the spectrum have received much interest \cite{Heiss_2005, Ribeiro_2007}, they can be viewed as QPT arising for excited states \cite{Caprio_2008}. In the semiclassical limit, this phenomena corresponds to a change in the topology of classical orbits and the appearance of hyperbolic points (HP). The non-analyticities involved are generically found to be logarithmic \cite{Heiss_2005, Ribeiro_2007, Caprio_2008}. 

\begin{figure}[t]
  \centering
\includegraphics[width=\columnwidth]{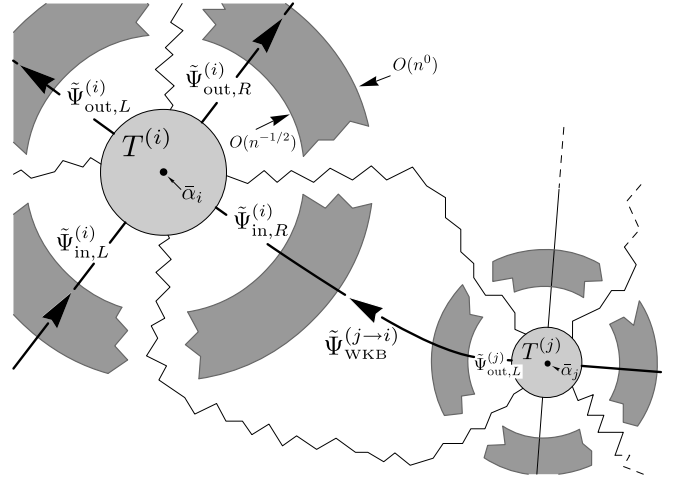}
\caption{ Phase space portrait of a classical trajectory $\mathcal{C}_0$ (full lines) describing a critical orbit that passing through two HP $\bar\alpha_i$ and $\bar\alpha_j$.  For $ O(n^{-1/2} ) < |\bar\alpha - \bar\alpha_i| < O(n^{0} )$ both, the linearized solutions around HP point and the WKB solutions, coexist (dark gray region), permitting to identify both asymptotic behaviors. The ``in'' and ``out'' solutions are connected via the $T^{(i)}$ matrices. Branch cuts of the WKB solutions are displayed as broken lines.}
  \label{fig:INandOUT}
  \end{figure}

We derive and test numerically spectral analytical expressions for two different situations  arising as a LMG-like model (LMG plus a cubic term) where hyperbolic trajectories, homoclinic and heteroclinic, exist.
Finally, we compute, both analytically and numerically, the matrix elements of observables between states near critical energies. We  discuss their semi-classical behavior hoping to clarify critical phenomena arising at the so-called excited states QPT.
%
%
%
%
\\
{\bf B-S Quantization and WKB. }
The non-normalized spin coherent states \cite{Perelomov_1986} for a $su(2)$ representation of dimension $2s+1$ are defined by $\ket{\alpha}=e^{ \alpha S_+}\ket{s,-s}$, $\bar\alpha \in \mathbb{C}$, with $s$, integer or half integer, being the total spin (in the following we set $n = 2s$). They form an over-complete basis with a resolution of the identity given by
$\int {\rm d}\mu \frac{\ket{\alpha } \bra{\alpha }}{\braket{\alpha}{\alpha}} = 1$,
where ${\rm d}\mu = \frac{{\rm d}\text{Re}(\alpha) {\rm d} \text{Im}(\alpha) }{\pi } \frac{n + 1}{\left(1+  \bar{\alpha }\alpha   \right)^{ 2 }} $ and $\braket{\alpha}{\alpha} = (1 + \bar\alpha \alpha)^{n}$.
In the coherent states basis, the $su(2)$ generators ($S_\pm=S_x \pm \rmi S_y$) act as differential operators 
%
%
\begin{eqnarray}
\label{eq:1}
S_+= n \bar\alpha-\bar\alpha^2 \partial_{\bar\alpha}; \ \ \
S_-= \partial_{\bar\alpha}; \ \ \
S_z= - \frac{n}{2}+ {\bar\alpha} \, \partial_{\bar\alpha},
\end{eqnarray}
%
%
on the space of polynomial functions $\Psi(\bar\alpha)= \braket{\alpha}{\Psi} $ of degree $n$. 
To a generic operator 
%
%
\begin{eqnarray}
\label{eq:2}
\hat{ A } =     \sum_{i} p_i( \bar\alpha ) \left( \frac{\partial_{\bar\alpha }}{n} \right)^i,
\end{eqnarray}
%
%
where the $p_i$'s are polynomials in $\bar\alpha$,
is associated a function (symbol) $\mathcal{ A }(\bar\alpha,\zeta ) = \sum_{i} p_i( \bar\alpha ) \zeta^i$.
In the framework of the WKB approximation, eigenstates of an Hermitian operator, $ \hat{ H } \Psi(\bar\alpha) =  \varepsilon \Psi(\bar\alpha) $, are obtained setting $\Psi(\bar\alpha) = e^{n \int_{\bar\alpha_0}^{\bar\alpha} G(\bar\alpha) \rm d \bar\alpha }$ and solving perturbatively the Riccati-like equation for $G$
%
%
\begin{eqnarray}
\label{eq:4}
\mathcal{H}\left[ \bar\alpha, G(\bar\alpha) + n^{-1}\partial_{\bar\alpha } \right] = \varepsilon,
\end{eqnarray}
%
%
in powers of the semi-classical parameter $n^{-1}$, setting 
%
%
\begin{eqnarray}
\label{eq:4_b}
\mathcal{ H } = \sum_{i=0}^{\infty} n^{-i} \mathcal{ H }_i\,; \ 
 G  = \sum_{i=0}^{\infty} n^{-i}   G_i \,; \ 
 \varepsilon  = \sum_{i=0}^{\infty} n^{-i}  \varepsilon_i.
\end{eqnarray}
%
%
The result is the WKB solution \cite{Kurchan_1989}, given in terms of ${\cal H}_i[\bar\alpha,G_0(\bar\alpha)]$,
%
%
\begin{eqnarray}
\label{eq:5}
\Psi_{\text{\tiny WKB}}(\bar\alpha) & =&  \sqrt{\frac{\partial_\zeta {\cal H}_0|_{\bar\alpha_0}}{\partial_\zeta {\cal H}_0}}e^{n \int_{\bar\alpha_0}^{\bar\alpha} {\rm d} \bar\alpha' \left[  G_0 +  \frac{1}{n} \frac{ \varepsilon_1 -{\cal H}_1+\frac{1}{2}\partial_\zeta \partial_{\bar\alpha} {\cal H}_0 }{\partial_\zeta {\cal H}_0 }  \right] }  \nonumber \\
& & \times [ 1+ O(n^{-1}) ].
\end{eqnarray}
%
%
Quantization of the energies is obtained by imposing that $\Psi(\bar\alpha)$ is a univaluated  function of $\bar\alpha \in \mathbb{C}$, implying that
%
%
\begin{eqnarray}
\label{eq:6}
\mathcal{I}_\gamma \equiv - \frac{1}{2 \pi \rmi} \oint_{\gamma} {\rm d} \bar\alpha \, G(\bar\alpha) = \frac{k}{n},
\end{eqnarray}
%
%
with $k\in \mathbb N$, for all closed path $\gamma$. In the semi-classical limit the probability amplitude $\braket{\alpha}{\alpha}^{-1} |\Psi(\bar\alpha)|^2$ of finding the system in the coherent state $\ket{\alpha}$ is exponentially localized on the classical trajectory  $\mathcal{C}_0 = \{ \bar \alpha: \  {\cal H}_0 \left( \bar \alpha, \frac{\alpha}{ 1+ \bar\alpha  \alpha}  \right) = \varepsilon_0 \}$, along which $G_0|_{\mathcal{C}_0} = \frac{\alpha}{1+\bar\alpha \alpha}$. Moreover, if $\mathcal{C}_0$ contains no HP, the WKB solution (\ref{eq:5}) is an analytic function of $\bar\alpha$ in its vicinity. In this case Eq.~(\ref{eq:6}) can be explicitly computed by choosing $\gamma = \mathcal{C}_0$ and using the semi-classical expansion of $G$. The result is the Bohr-Sommerfeld quantization condition for a spin system \cite{Shankar_1980,Kurchan_1989,Garg_2004}:
$
\mathcal{I}_0 + n^{-1} \mathcal{I}_1 + O(n^{-2}) = n^{-1} k;
$
where $
\mathcal{I}_0 = -\frac{1}{2 \pi \rmi} \oint_{ \mathcal{C}_0 } {\rm d} \bar\alpha \frac{ \alpha }{1+ \bar\alpha  \alpha } =  \int_{ \Sigma } \omega$
is the classical action obtained by integrated the symplectic 2-form $ \omega = \frac{1}{2 \pi \rmi} (1+ \bar\alpha \alpha )^{-2} \rm d \bar \alpha \land \rm d \alpha $ over the interior of the classical trajectory  $\Sigma$, and 
$
\mathcal{I}_1 = \frac{1}{2} - \frac{1}{2 \pi \rmi} \oint_{ \mathcal{C}_0 } {\rm d} \bar\alpha \frac{ \varepsilon_1 -{\cal H}_1+\frac{1}{2}\partial_\zeta \partial_{\bar\alpha} {\cal H}_0 }{\partial_\zeta {\cal H}_0 }$.
%
%
%
%
\\
\\
{\bf Quantization near HP.}
If an HP exits along classical trajectory, i.e. $\partial_\zeta {\cal H}_0 = 0 $ for some $\bar\alpha_i \in \mathcal{C}_0$, $\mathcal{I}_1$ diverges and the quantization condition has to be modified for energies of order $n^{-1}$ around the critical energy $\varepsilon_0 = \varepsilon_c$.
Near such points, setting $\bar \beta = \bar \alpha - \bar \alpha_i$, $\cal H$ can be linearized and brought to the form
%
%
\begin{eqnarray}
\label{eq:10}
\tilde {\cal H} ( \bar \beta, \zeta)  - \varepsilon =  \tau_2 \, \zeta^2 + \tau_0 \, \bar\beta^2 + \frac{\tau_{00}  - \varepsilon_1 }{n}  \, +\,  O( |\bar\beta|^{3} ),
\end{eqnarray}
%
%
by a simple transformation $\Psi(\bar{\alpha }) =  e^{ n \, p(\bar{\beta } )  }   \tilde \Psi (\bar{\beta } )$, where $p$ is a second order polynomial of $\bar\beta$. The constants $\tau_k$ depend on the parameter of the Hamiltonian around the HP. The solutions of $\left[ \tilde {\cal H} ( \bar \beta, n^{-1}\partial_{\bar \beta})  - \varepsilon \right] \tilde \Psi (\bar{\beta } ) = 0 $ are given explicitly in the form of Parabolic Cylindrical functions \cite{Gradshteyn_1980}. Let us consider the following linear combinations of these two independent solutions, having a well defined behavior when $ |\bar \beta| \, n^{1/2}  \to  \infty$, for $\bar \beta$ in a vicinity of $\mathcal{C}_0$ (see Fig.~\ref{fig:INandOUT} for the directions along which each limit is taken),
%
%
\begin{eqnarray}
\label{eq:11}
\left.
\begin{array}{l}
\tilde \Psi_{\text{out},R} (\bar{\beta } ) \\
\tilde \Psi_{\text{in},L} (\bar{\beta } )
\end{array}
\right\}
&\to&
 e^{-  \rmi  n \rho ^2   \bar{\beta }^2  } \bar{\beta }^{-\frac{1}{2} + \rmi \eta } \left[ 1 + O( |\bar\beta|^{-1} n^{-1/2} ) \right], \nonumber  \\
\left.
\begin{array}{l}
\tilde \Psi_{\text{out},L} (\bar{\beta } ) \\
\tilde \Psi_{\text{in},R} (\bar{\beta } )
\end{array}
\right\}
&\to&
 e^{  \rmi  n \rho ^2  \bar{\beta }^2  } \bar{\beta }^{  -\frac{1}{2} - \rmi \eta } \left[ 1 + O( |\bar\beta|^{-1} n^{-1/2} ) \right], \nonumber
\end{eqnarray}
%
%
where $\rho = \left| \frac{\tau _0}{4 \tau _2} \right|^{1/4}  $ and $\eta = \frac{\varepsilon_1 -\tau _{00}}{4 \rho ^2 \tau _2} $. 
%
%
%
%
Being solutions of a second order differential equation, these four functions are obviously not independent. The explicit form of the Parabolic Cylindrical functions provides a ``connection'' between different asymptotic regions
%
%
\begin{eqnarray}
\label{eq:12}
\left(
\begin{array}{l}
\tilde \Psi_{\text{out},L}  \\
\tilde \Psi_{\text{out},R}
\end{array}
\right)
=
T
\left(
\begin{array}{l}
\tilde \Psi_{\text{in},R}   \\
\tilde \Psi_{\text{in},L}  
\end{array}
\right),\\
T = \left(
\begin{array}{ll}
 1 & -\bar{c} \\
 c & -e^{-2 \pi  \eta }
\end{array} 
\right)
+ O(n^{-1}),
\end{eqnarray}
%
%
with $ c = \sqrt{\frac{2 e^{-\pi  \eta } }{\pi }} \cosh (\pi  \eta ) e^{-\rmi \left[ \eta  \log \left(4 n \rho ^2\right)+\frac{\pi }{2}\right]} \Gamma \left(\rmi \eta +\frac{1}{2}\right) $.

\begin{figure}
  \centering
\includegraphics[width= 8.5 cm ]{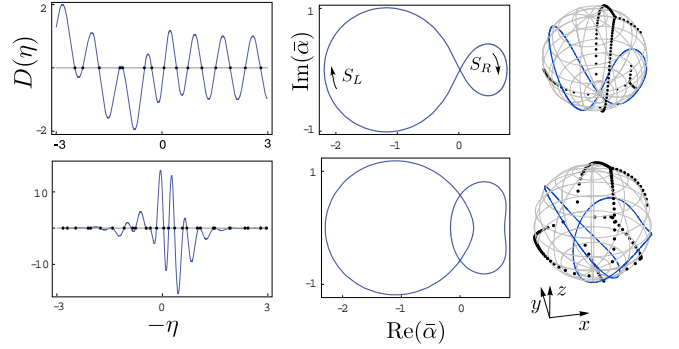}
\caption{ 
Homoclinic Case (Up): $h = 1; \gamma_x = 4; \gamma_y = 1/4; \mu = 5$. 
Heteroclinic Case (Down): $h = 1; \gamma_x = 5; \gamma_y = 2; \mu = 6$.
Left: Comparison between the zeros of $D$ (blue line) and the eigenvalues of (\ref{eq:17}) computed numerically (black dots) for $ n = 500 $. 
We define the renormalized energy $\eta = -\frac{h+\varepsilon_1 }{2 \sqrt{ (\gamma _x-h) (h-\gamma_y)}}
$;  $ \eta=\frac{\eta_1 + \eta_2}{2} = \frac{- (\lambda +\gamma _y) \sqrt{\gamma _y}}{2 \sqrt{\left(\gamma _x-\gamma _y\right) \left(\gamma_y^2-h^2\right)}}$ respectively for the homoclinic and heteroclinic cases.
Middle: Stereographic projection of the critical classical orbit. Right: Critical orbit in Riemann Sphere, the zeroes of $\Psi(\bar\alpha)$ (black dots) are plotted for $n = 120$, in the semiclassical limit they condense in branch cuts of $G_0$ \cite{Ribeiro_2007, Nonnenmacher_1997}. }
  \label{fig:HeteroAndHomo}
  \end{figure}

Constrains of the type (\ref{eq:12}) give a set of local relations between the ``in'' and ``out'' basis. A set of non-local relations is obtained by identifying the asymptotics of WKB solutions (see Fig.~\ref{fig:INandOUT}), leading to
$\tilde \Psi_{\text{out},L} = e^{ 2 \pi \rmi n S(\bar\alpha_i , \bar\alpha_j ) }  \tilde \Psi_{\text{in},R}$,
where $S(\bar\alpha_i , \bar\alpha_j )$ is regularized action integral given in Table~\ref{tb:Table1},  $\nu_j = ( \pm \rmi \eta_j-\frac{1}{2})$ depending on the side $R/L$ and $j$ indexing HP. $\ln(x)$ is defined as having a branch cut along the negative real axes. $\sigma_k = 0,\pm 1$: $0$ if the classical orbit does not cut the branch-cut of $\ln (\bar\alpha - \bar\alpha_k )$ and $\pm1$ if it cuts it in the up-down or down-up directions respectively. 
Summarizing the local and non-local basis relations:
$ \bf \Psi_{\text{out}} =  T  \Psi_{\text{in} }$ and $ \bf \Psi_{\text{out}} = \Gamma  \Psi_{\text{in} }$;
where $ \bf \Psi_{\text{out}}$ and $\bf \Psi_{\text{in}}$ are column vectors collecting the ``in'' and ``out'' solutions for each HP $(i)$, $\bf T$ and $\bf  \Gamma$ are  matrices, the first coupling states with the same $(i)$ and the second coupling sates with $(i)$ and $(j)$ linked by the classical trajectory.
Quantization is obtained by imposing the compatibility relation $D = \det ( {\bf  T - \Gamma} ) = 0$.
\begin{widetext} 
\begin{table}[h]
\begin{tabular}{cc}
 {\small Heteroclinic} & 
$\begin{array}{l}
2\pi \rmi S(\bar\alpha_i , \bar\alpha_j )  = 2\pi \rmi S_{i,j} 
+ \frac{1}{n}  \Big\{    \nu_j \ln[ (-1)^{\sigma_j}(\bar\alpha_i - \bar\alpha_j )]  
- \nu_i \ln [(-1)^{\sigma_i}(\bar\alpha_j - \bar\alpha_i )]  
 + \sigma_i \rmi \pi \nu_i  - \sigma_j \rmi \pi \nu_j \Big\} ;
\\
2\pi \rmi S_{i,j} = \int_{\bar\alpha_j}^{\bar\alpha_i} \frac{\alpha}{1+ \bar\alpha \alpha} { \rm d}\bar\alpha - \frac{1}{n} \int_{\bar\alpha_j}^{\bar\alpha_i}  
 \partial_{\bar\alpha}\left[  (\bar\alpha - \bar\alpha_i )(\bar\alpha - \bar\alpha_j )G_1 \right]   
\frac{\ln [(-1)^{\sigma_i}(\bar\alpha - \bar\alpha_i )] - \ln [(-1)^{\sigma_j}(\bar\alpha - \bar\alpha_j )]}{\bar\alpha_i - \bar\alpha_j }  { \rm d}\bar\alpha;
\end{array}$
\\ \hline
 {\small Homoclinic}   & 
$\begin{array}{l}
2\pi \rmi S(\bar\alpha_i , \bar\alpha_i )  =  2\pi \rmi S_i 
+ \frac{\rmi \pi \sigma}{n};  
\\
2\pi \rmi S_i = \int_{\bar\alpha_j}^{\bar\alpha_i} \frac{\alpha}{1+ \bar\alpha \alpha} { \rm d}\bar\alpha 
- \frac{1}{n}  
\int_{\bar\alpha_j}^{\bar\alpha_i}   \ln [(-1)^{\sigma}(\bar\alpha - \bar\alpha_i )]    \partial_{\bar\alpha}\left[  (\bar\alpha - \bar\alpha_i ) G_1 \right] { \rm d}\bar\alpha ;
\end{array}$
\end{tabular}\caption{Regularized Action Integrals}\label{tb:Table1}
\end{table}
\end{widetext}
We now apply the general method presented above to a particular spin Hamiltonian 
%
%
\begin{eqnarray}
\label{eq:17}
\hat H = \frac{2}{n} \left(  h S_z - \frac{\gamma_x S_x^2  + \gamma_y S_y^2 }{n} + \mu  \frac{S_x^3  }{n^2}                \right).
\end{eqnarray}
%
%
The Lipkin-Meshkov-Glick (LMG) model \cite{Lipkin_1965} is obtained from Eq.~(\ref{eq:17}) setting $\mu=0$.
The cubic term in Eq.~(\ref{eq:17}) is added to provide asymmetric orbits in order to test the quantization relations in a case as generic as possible. For the LMG model a detailed analysis of the phase space and the characterization of the critical points can be found in \cite{Castanos_2006, Ribeiro_2007, Ribeiro_2008}. For small values of $\mu$ the phase diagram presented in \cite{Ribeiro_2007} is kept invariant. In particular the system conserves a homoclinic HP at $\alpha=0$ for $\varepsilon_c = -|h|$ when $\gamma_x > |h| < |\gamma_y|$ and a heteroclinic caustic joining two HP for $\gamma_x > \gamma_y > |h|$ corresponding to $\varepsilon_c = - \frac{h^2 + \gamma_y^2}{ 2 \gamma_y}$. 
For the homoclinic case, one obtains
$
D = -\frac{\cos \left[\pi  n  \left(S_L+S_R\right)\right]}{\sqrt{1+e^{-2 \pi  \eta }}} 
-\sin \big\{ \arg \big[ \Gamma (1/2-i
   \eta ) \big] +\eta  \log \left(4 \rho ^2 n \right)+\pi  n  \left(S_R-S_L\right)\big\}
$, as in the case of the Schr\"odinger representation \cite{Verdiere_1999}, where $S_{R/L}$ are given by $S_i$ in Table~\ref{tb:Table1} (directions of integration are given in Fig.~\ref{fig:HeteroAndHomo}). For the heteroclinic case the quantization condition is rather lengthy and will be given elsewhere \cite{Paul_2008}. The comparison of the semi-classical quantization conditions with numeric diagonalization of the Hamiltonian using a matricial representation of the spin operators is given in Fig.~\ref{fig:HeteroAndHomo}. In both cases the agreement between the numeric energies and the points where $D=0$ is remarkable, for the heteroclinic case one can see that the matching becomes worst as the modulus of the renormalized energy $\eta$ increases.
%
%
%
%
%
%
%
%
\\
{\bf Matrix Elements. }  
In the semiclassical limit, the normalized matrix elements $f^{A}_k(\varepsilon^{(m)}) = \frac{\bra{ \Psi_{m+k} }\hat A \ket{ \Psi_{m} }}{ \sqrt{\braket{ \Psi_{m+k} }{ \Psi_{m+k} }\braket{ \Psi_{m} }{ \Psi_{m} } }}$, of an observable $\hat A$ computed between eigenstates of an Hermitian operator $H$ (with the energies $\varepsilon^{(m)}$ and $\varepsilon^{(m+k)}$ ), are known to be simply given as the amplitude of the  $k$-th Fourier mode of the observable symbol $\rm A$, evaluated along the classical orbit of energy $\varepsilon^{(m)}$ \cite{Paul_1993},
$
f^{A}_k(\varepsilon^{(m)}) = \frac{1}{T}\int_{-T/2}^{T/2} {\rm d}t \, {\rm e}^{ \rmi k \frac{2\pi}{T}} {\cal A}[\bar\alpha(t),\zeta(t)],
$
where $T$ is the period of the classical orbit and the flow equations are given by: $ \partial_t \bar\alpha(t) = - {\rm i}\, \partial_\zeta {\cal H}(\bar\alpha, \zeta); \ \ \partial_t \zeta(t) =  {\rm i}\, \partial_{\bar\alpha} {\cal H}(\bar\alpha, \zeta)$ [$\zeta$ is the variable conjugated to $\bar\alpha$, for the spin case $\zeta = \alpha (1+\bar\alpha \alpha)^{-1}$]. This result stands for regular orbits and can be obtained considering the action-angle variables. Since $f$ is the Fourier transform of a analytic function the matrix elements vanish exponentially with increasing $k$. This is a generalization of the result early obtained by Heisenberg for the harmonic oscillator case.

\begin{figure}[ht]
  \centering
\includegraphics[width= 8 cm ]{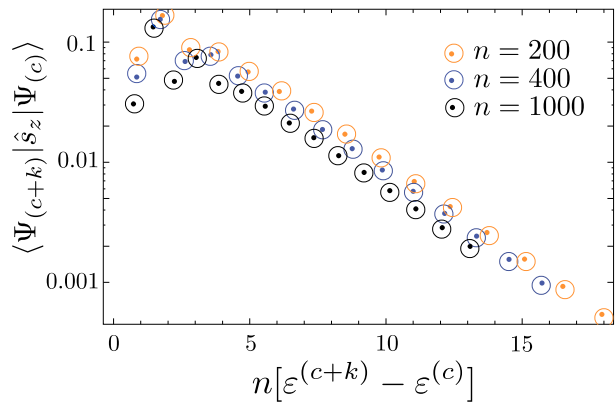}
\caption{ Matrix elements of the operator $\hat s_z = \frac{\hat S_z}{s}$ between two states near the critical energy. $\varepsilon^{(c)}$ is chosen to be the energy closest to the critical classical energy $\varepsilon_c$. The agreement of the numerical data (dots) with the predictions of Eq.~(\ref{eq:22}) (circles) gets better for $n$ big. The logarithmic downward shift as $n$ increases is due to the fact that $\mu(i) \sim \ln^{-1} n$.
 }
  \label{fig:MatrixEll}
  \end{figure}

For singular orbits containing HP the period $T$ diverges, moreover no action-angle variables can be defined. Nevertheless it is still possible to estimate such matrix elements by analyzing local and global properties of the critical eigenstates \cite{Cary_1993}. Let us use the resolution of the identity in order to write matrix elements as integrals over $\Sigma_{i}$, a domain of size $O( n^{-1} \ln n)$ around the HP $\bar\alpha_i$, and $\Sigma_{i,j}$, a domain of order $n^{-1}$ around ${\cal C}_0$. Within these two sets of domains the eigenstates are given, respectively, by special functions and WKB approximation, 
%
%
\begin{multline}
\label{eq:20} 
\bra{ \Psi_{m+k} }\hat A \ket{ \Psi_{m} } = \\
= \left[ \sum_{(i)\to(j)} \int_{\Sigma_{i,j}}  + \sum_{(i)} \int_{\Sigma_{i}}  \right] \frac{\braket{\Psi_{m+k}}{\alpha}\bra{\alpha}\hat A \ket{\Psi_m}}{\braket{\alpha}{\alpha}}  {\rm d}\mu  \\
= \sum_{(i)\to(j)} g^{A}_{i \to j}[ n (\varepsilon^{(m+k)} -\varepsilon^{(m)})] + \delta_{k,0} \sum_{(i)} {\cal A}(\bar \alpha_i, \zeta_i)  \mu(i).  \nonumber
\end{multline}
%
%
The last equality follows from considering the symbol ${\cal A}$ constant on the domain $\Sigma_{i}$, by orthogonality of the eigenstates this term is nonzero only for $k=0$ where it gives the norm of the eigenstate inside the domain, $\mu(i)$. The regular functions $g^{A}_{i \to j}(\omega)= \int_{-\infty}^{\infty}{\rm d}t\, {\cal A}(t) {\rm e}^{{\rm i} t \omega }$ are computed using the flow equations on the branch $i\to j$. Since $\mu(i) \propto \ln n$, we obtain at leading order,
%
%
\begin{eqnarray}
\label{eq:21} 
f_{k=0}^A(\varepsilon^{(m)}) &=&  \frac{\sum_{(i)} {\cal A}(\bar \alpha_i, \zeta_i)  \mu(i)}{\sum_{(i)}   \mu(i)}, \\
\label{eq:22} 
f_{k\neq0}^A(\varepsilon^{(m)}) &=&  \frac{\sum_{(i)\to(j)} g^{A}_{i \to j}[n (\varepsilon^{(m+k)} -\varepsilon^{(m)})] }{\sum_{(i)}   \mu(i)}.
\end{eqnarray}
%
%
Diagonal matrix elements (mean values of observables) are thus given as a sum of ponderate weights of the different HP and depend on local properties of eigenstates near this points. On the contrary, non-diagonal elements are given by the global properties of the classical orbit. Since $g^{A}$ is analytic, the matrix elements will decay exponentially as the energy difference increases, however, near the critical energy the mean energy spacing is of order $n(\varepsilon^{(m+k)} -\varepsilon^{(m)} )\propto k \ln^{-1} n$, meaning that the exponential decay in $k$ becomes slower with increasing $n$ (see Fig.~\ref{fig:MatrixEll}). 
For an observable with $\cal A$ vanishing at the HP the amplitude of all matrix elements vanishes as $O(\ln^{-1} n)$, for fixed $k$ (Fig.~\ref{fig:MatrixEll}) \cite{Paul_2008}. This has a simple semi-classical explanation.
In the critical case the volume of the phase-space corresponding to an energy band of  order $n^{-1}$ around $\varepsilon_c$ is $O(n^{-1})$ for regions of type $\Sigma_{i,j}$ and $O(n^{-1} \ln{n})$ for $\Sigma_i$. 
However, for $\cal A$ vanishing at the HP, the relevant regions to compute the matrix elements are $\Sigma_{i,j}$ which, by Heisenberg inequalities, 
can carry only a finite number of states $O(n^0)$ and not  the total $O(\ln n)$ eigenstates.
The only way of conciliating these two facts is to take a quantized observable described by an $O(\ln{n})\times O(\ln{n})$ matrix whose elements vanish in the classical limit.
%
%
%
%
%
%
%
%
\\
{\bf Conclusion.}
We have presented a method for computing semi-classical spectra associated to any number of heteroclinic junctions. 
Not only the expected average spacing $\sim \ln^{-1}n$ is  observed, but an algebraic relation is derived for eigenvalues near the critical energy. 
The method is fully general and applies to any $su(2)$ Hamiltonian. In order to test it in full generality we have added a cubic term to the standard LMG model breaking the quadratic underlying symmetry.
The agreement with numerics is remarkable, especially considering the fact that the formulas
are algebraically quite heavy in the case of two hyperbolic fixed points linked by heteroclininc junctions. 
We have also computed the matrix elements of observables, and show that their semi-classical behavior is universal, and different from the one in the regular situation. 
Moreover we have given a physical argument for the logarithmic vanishing of these matrix elements in the classical limit.
\\
\acknowledgments
We are grateful to  R. Mosseri, J. Vidal for fruitful and stimulating discussions and to P. Vieira for a careful reading of the manuscript. 
PR was partially supported by FCT and EU FEDER through POCTI and POCI, namely via QuantLog POCI/MAT/55796/2004 Project of CLC-DM-IST, SQIG-IT and grant SFRH/BD/16182/2004/2ZB5.

\begin{thebibliography}{21}
\expandafter\ifx\csname natexlab\endcsname\relax\def\natexlab#1{#1}\fi
\expandafter\ifx\csname bibnamefont\endcsname\relax
  \def\bibnamefont#1{#1}\fi
\expandafter\ifx\csname bibfnamefont\endcsname\relax
  \def\bibfnamefont#1{#1}\fi
\expandafter\ifx\csname citenamefont\endcsname\relax
  \def\citenamefont#1{#1}\fi
\expandafter\ifx\csname url\endcsname\relax
  \def\url#1{\texttt{#1}}\fi
\expandafter\ifx\csname urlprefix\endcsname\relax\def\urlprefix{URL }\fi
\providecommand{\bibinfo}[2]{#2}
\providecommand{\eprint}[2][]{\url{#2}}

\bibitem[{\citenamefont{Connor}(1969)}]{Connor_1969}
\bibinfo{author}{\bibfnamefont{J.~N.~L.} \bibnamefont{Connor}},
  \bibinfo{journal}{Chem. Phys. Lett.} \textbf{\bibinfo{volume}{4}},
  \bibinfo{pages}{419} (\bibinfo{year}{1969}).

\bibitem[{\citenamefont{{Colin de Verdi{\`e}re} and
  Parisse}(1999)}]{Verdiere_1999}
\bibinfo{author}{\bibfnamefont{Y.}~\bibnamefont{{Colin de Verdi{\`e}re}}}
  \bibnamefont{and} \bibinfo{author}{\bibfnamefont{B.}~\bibnamefont{Parisse}},
  \bibinfo{journal}{Commun. Math. Phys.} \textbf{\bibinfo{volume}{205}},
  \bibinfo{pages}{459} (\bibinfo{year}{1999}).

\bibitem[{\citenamefont{Lipkin et~al.}(1965)\citenamefont{Lipkin, Meshkov, and
  Glick}}]{Lipkin_1965}
\bibinfo{author}{\bibfnamefont{H.~J.} \bibnamefont{Lipkin}},
  \bibinfo{author}{\bibfnamefont{N.}~\bibnamefont{Meshkov}}, \bibnamefont{and}
  \bibinfo{author}{\bibfnamefont{A.~J.} \bibnamefont{Glick}},
  \bibinfo{journal}{Nucl. Phys.} \textbf{\bibinfo{volume}{62}},
  \bibinfo{pages}{188} (\bibinfo{year}{1965}).

\bibitem[{\citenamefont{Garanin et~al.}(1998)\citenamefont{Garanin, X.,
  Hidalgo, and Chudnovsky}}]{Garanin_1998}
\bibinfo{author}{\bibfnamefont{D.~A.} \bibnamefont{Garanin}},
  \bibinfo{author}{\bibnamefont{X.}},
  \bibinfo{author}{\bibfnamefont{M.}~\bibnamefont{Hidalgo}}, \bibnamefont{and}
  \bibinfo{author}{\bibfnamefont{E.~M.} \bibnamefont{Chudnovsky}},
  \bibinfo{journal}{Phys. Rev. B} \textbf{\bibinfo{volume}{57}},
  \bibinfo{pages}{13639} (\bibinfo{year}{1998}).

\bibitem[{\citenamefont{Ulyanov and Zaslavskii}(1992)}]{Ulyanov_1992}
\bibinfo{author}{\bibfnamefont{V.~V.} \bibnamefont{Ulyanov}} \bibnamefont{and}
  \bibinfo{author}{\bibfnamefont{O.~B.} \bibnamefont{Zaslavskii}},
  \bibinfo{journal}{Phys. Rep.} \textbf{\bibinfo{volume}{216}},
  \bibinfo{pages}{179} (\bibinfo{year}{1992}).

\bibitem[{\citenamefont{Vidal et~al.}(2004)\citenamefont{Vidal, Palacios, and
  Mosseri}}]{Vidal_2004_b}
\bibinfo{author}{\bibfnamefont{J.}~\bibnamefont{Vidal}},
  \bibinfo{author}{\bibfnamefont{G.}~\bibnamefont{Palacios}}, \bibnamefont{and}
  \bibinfo{author}{\bibfnamefont{R.}~\bibnamefont{Mosseri}},
  \bibinfo{journal}{Phys. Rev. A} \textbf{\bibinfo{volume}{69}},
  \bibinfo{pages}{022107} (\bibinfo{year}{2004}).

\bibitem[{\citenamefont{Sachdev}(1999)}]{Sachdev_1999}
\bibinfo{author}{\bibfnamefont{S.}~\bibnamefont{Sachdev}},
  \emph{\bibinfo{title}{Quantum Phase Transitions}}
  (\bibinfo{publisher}{Cambridge University Press}, \bibinfo{year}{1999}).

\bibitem[{\citenamefont{Heiss et~al.}(2005)\citenamefont{Heiss, Scholtz, and
  Geyer}}]{Heiss_2005}
\bibinfo{author}{\bibfnamefont{W.~D.} \bibnamefont{Heiss}},
  \bibinfo{author}{\bibfnamefont{F.~G.} \bibnamefont{Scholtz}},
  \bibnamefont{and} \bibinfo{author}{\bibfnamefont{H.~B.} \bibnamefont{Geyer}},
  \bibinfo{journal}{J. Phys. A} \textbf{\bibinfo{volume}{38}},
  \bibinfo{pages}{1843} (\bibinfo{year}{2005}).

\bibitem[{\citenamefont{Ribeiro et~al.}(2007)\citenamefont{Ribeiro, Vidal, and
  Mosseri}}]{Ribeiro_2007}
\bibinfo{author}{\bibfnamefont{P.}~\bibnamefont{Ribeiro}},
  \bibinfo{author}{\bibfnamefont{J.}~\bibnamefont{Vidal}}, \bibnamefont{and}
  \bibinfo{author}{\bibfnamefont{R.}~\bibnamefont{Mosseri}},
  \bibinfo{journal}{Phys. Rev. Lett.} \textbf{\bibinfo{volume}{99}},
  \bibinfo{eid}{050402} (\bibinfo{year}{2007}).

\bibitem[{\citenamefont{Caprio et~al.}(2008)\citenamefont{Caprio, Cejnar, and
  Iachello}}]{Caprio_2008}
\bibinfo{author}{\bibfnamefont{M.}~\bibnamefont{Caprio}},
  \bibinfo{author}{\bibfnamefont{P.}~\bibnamefont{Cejnar}}, \bibnamefont{and}
  \bibinfo{author}{\bibfnamefont{F.}~\bibnamefont{Iachello}},
  \bibinfo{journal}{Ann. Phys.} \textbf{\bibinfo{volume}{323}},
  \bibinfo{pages}{1106} (\bibinfo{year}{2008}).

\bibitem[{\citenamefont{Perelomov}(1986)}]{Perelomov_1986}
\bibinfo{author}{\bibfnamefont{A.}~\bibnamefont{Perelomov}},
  \emph{\bibinfo{title}{Generalized Coherent States and Their Applications}}
  (\bibinfo{publisher}{Springer}, \bibinfo{year}{1986}).

\bibitem[{\citenamefont{Kurchan et~al.}(1989)\citenamefont{Kurchan, Leboeuf,
  and Saraceno}}]{Kurchan_1989}
\bibinfo{author}{\bibfnamefont{J.}~\bibnamefont{Kurchan}},
  \bibinfo{author}{\bibfnamefont{P.}~\bibnamefont{Leboeuf}}, \bibnamefont{and}
  \bibinfo{author}{\bibfnamefont{M.}~\bibnamefont{Saraceno}},
  \bibinfo{journal}{Phys. Rev. A} \textbf{\bibinfo{volume}{40}},
  \bibinfo{pages}{6800} (\bibinfo{year}{1989}).

\bibitem[{\citenamefont{Shankar}(1980)}]{Shankar_1980}
\bibinfo{author}{\bibfnamefont{R.}~\bibnamefont{Shankar}},
  \bibinfo{journal}{Phys. Rev. Lett.} \textbf{\bibinfo{volume}{45}},
  \bibinfo{pages}{1088} (\bibinfo{year}{1980}).

\bibitem[{\citenamefont{Garg and Stone}(2004)}]{Garg_2004}
\bibinfo{author}{\bibfnamefont{A.}~\bibnamefont{Garg}} \bibnamefont{and}
  \bibinfo{author}{\bibfnamefont{M.}~\bibnamefont{Stone}},
  \bibinfo{journal}{Phys. Rev. Lett.} \textbf{\bibinfo{volume}{92}},
  \bibinfo{pages}{010401} (\bibinfo{year}{2004}).

\bibitem[{\citenamefont{Gradshteyn and Ryzhik}(1980)}]{Gradshteyn_1980}
\bibinfo{author}{\bibfnamefont{I.}~\bibnamefont{Gradshteyn}} \bibnamefont{and}
  \bibinfo{author}{\bibfnamefont{I.}~\bibnamefont{Ryzhik}},
  \emph{\bibinfo{title}{Table of Integrals, Series and Product}}
  (\bibinfo{publisher}{Academic Press}, \bibinfo{address}{New-York},
  \bibinfo{year}{1980}).

\bibitem[{\citenamefont{Castanos et~al.}(2006)\citenamefont{Castanos,
  Lopez-Pena, Hirsch, and Lopez-Moreno}}]{Castanos_2006}
\bibinfo{author}{\bibfnamefont{O.}~\bibnamefont{Castanos}},
  \bibinfo{author}{\bibfnamefont{R.}~\bibnamefont{Lopez-Pena}},
  \bibinfo{author}{\bibfnamefont{J.~G.} \bibnamefont{Hirsch}},
  \bibnamefont{and}
  \bibinfo{author}{\bibfnamefont{E.}~\bibnamefont{Lopez-Moreno}},
  \bibinfo{journal}{Phys. Rev. B} \textbf{\bibinfo{volume}{74}},
  \bibinfo{pages}{104118} (\bibinfo{year}{2006}).

\bibitem[{\citenamefont{Ribeiro et~al.}()\citenamefont{Ribeiro, Vidal, and
  Mosseri}}]{Ribeiro_2008}
\bibinfo{author}{\bibfnamefont{P.}~\bibnamefont{Ribeiro}},
  \bibinfo{author}{\bibfnamefont{J.}~\bibnamefont{Vidal}}, \bibnamefont{and}
  \bibinfo{author}{\bibfnamefont{R.}~\bibnamefont{Mosseri}},
  \bibinfo{note}{arXiv:0805.4078}.

\bibitem[{\citenamefont{Paul and Ribeiro}(2008)}]{Paul_2008}
\bibinfo{author}{\bibfnamefont{T.}~\bibnamefont{Paul}} \bibnamefont{and}
  \bibinfo{author}{\bibfnamefont{P.}~\bibnamefont{Ribeiro}}
  (\bibinfo{year}{2008}), \bibinfo{note}{in preparation}.

\bibitem[{\citenamefont{Nonnenmacher and Voros}(1997)}]{Nonnenmacher_1997}
\bibinfo{author}{\bibfnamefont{S.}~\bibnamefont{Nonnenmacher}}
  \bibnamefont{and} \bibinfo{author}{\bibfnamefont{A.}~\bibnamefont{Voros}},
  \bibinfo{journal}{J. Phys. A} \textbf{\bibinfo{volume}{30}},
  \bibinfo{pages}{295} (\bibinfo{year}{1997}).

\bibitem[{\citenamefont{Paul and Uribe}(1993)}]{Paul_1993}
\bibinfo{author}{\bibfnamefont{T.}~\bibnamefont{Paul}} \bibnamefont{and}
  \bibinfo{author}{\bibfnamefont{A.}~\bibnamefont{Uribe}},
  \bibinfo{journal}{Ann. Inst. Henri Poincar{\'e} (A)}
  \textbf{\bibinfo{volume}{59}}, \bibinfo{pages}{357} (\bibinfo{year}{1993}).

\bibitem[{\citenamefont{Cary and Rusu}(1993)}]{Cary_1993}
\bibinfo{author}{\bibfnamefont{J.~R.} \bibnamefont{Cary}} \bibnamefont{and}
  \bibinfo{author}{\bibfnamefont{P.}~\bibnamefont{Rusu}},
  \bibinfo{journal}{Phys. Rev. A} \textbf{\bibinfo{volume}{47}},
  \bibinfo{pages}{2496} (\bibinfo{year}{1993}).

\end{thebibliography}
\end{document}